\pdfoutput=1

\documentclass[11pt]{article} 
\usepackage{authblk}

\usepackage[utf8]{inputenc} 
\usepackage{amsmath}
\usepackage{amssymb}
\usepackage{bm}
\usepackage{float}
\usepackage{multicol}

\usepackage{geometry} 
\usepackage{graphicx} 
\usepackage{subcaption}
\graphicspath{{./figs/}}
\DeclareGraphicsExtensions{.eps,.jpg}

\usepackage{booktabs} 
\usepackage{array} 
\usepackage{paralist} 
\usepackage{verbatim} 
\usepackage{url}						

\usepackage{fancyhdr} 
\usepackage{sectsty}

\usepackage[nottoc,notlof,notlot]{tocbibind} 
\usepackage[titles,subfigure]{tocloft} 

\title{Review of Algorithms for Compressive Sensing of Images}
\author[1]{Yoni Sher}
\affil[1]{School of Computer Science and Engineering, Hebrew University of Jerusalem, Jerusalem 91904, Israel}
\date{} 

\begin{document}
\maketitle

\begin{abstract}
	We provide a comprehensive review of classical algorithms for compressive sensing of images, focused on Total variation methods, with a view to application in LiDAR systems. Our primary focus is providing a full review for beginners in the field, as well as simulating the kind of noise found in real LiDAR systems. To this end, we provide an overview of the theoretical background, a brief discussion of various considerations that come in to play in compressive sensing, and a standardized comparison of off-the-shelf methods, intended as a quick-start guide to choosing algorithms for compressive sensing applications. 
\end{abstract}

\section{Introduction and theoretical background}
This paper is intended as a "how-to" guide for beginners in the field of compressive sensing, giving a broad introduction to the field and the classical algorithms available. The comparative section is written in the spirit of \cite{yang2011alternating, becker2011nesta} and others, however with the focus of running time of given algorithms "as provided", for the benefit of users without the resources to develop and deploy custom software packages. We focus on classical algorithms requiring no parallelization or special hardware support, suitable for generic workstation environments. \\
We begin with a short explanation of Compressive Sensing: the theoretical background and the assumptions it is based on. We then give a qualitative discussion of the major algorithms for sparse image reconstruction, from standard algorithms available with every scientific computation package to optimized special-purpose methods. Finally, We present the results of comprehensive simulation of selected algorithms under a variety of parameters and quantitative comparison of the different methods. 

\subsection{Sub-sampling theory}
Compressive Sensing is a method of reconstructing images from fewer measurements than the number of pixels in the image. It is based on two assumptions: that the image has a sparse representation in some basis, and that the measurement basis in incoherent with the basis in which the image is sparse \cite{candes2007sparsity}.  \\
Sparsity gives rise to the following argument: the Shannon entropy of an image with $ n $ pixels but only $ k \ll n $ non-zero elements is \cite{howland2014compressive} 
\begin{equation}\label{entropy}
	-\frac{k}{n}  \log{\frac{k}{n}} -\left(1-\frac{k}{n}\right)  \log\left(1-\frac{k}{n}\right) \approx \frac{k}{n} \log{\frac{k}{n}} \approx k \log \frac{n}{k} 
\end{equation}
We use this as a lower bound for the number of measurement required to reconstruct a signal. We will show that we can get very close to this bound. \\
Incoherence \cite{candes2008introduction,howland2014compressive}: Two bases are mutually incoherent if the dot product of any vector from one with any vector from the other is not too large: If $\mathbf{\phi,\psi}$  are basis of dimension $n$, the mutual incoherence is:
\begin{equation}\label{mutual_incoherence}
\mu(\mathbf{\phi,\psi}) = \sqrt{n} \max_{i,j} \|\langle\phi_i, \psi_j \rangle\|
\end{equation}
Natural images tend to be sparse in frequency bases: Fourier or Wavelet, for example. Random measurement bases will be approximately $ \sqrt{2 \log n} $ incoherent with any base with high probability (and therefore suitable for compressed sensing).
\subsection{Reconstruction theory: }
\label{recon_theory}
Let $ x $ some signal, and $ \boldsymbol{\phi} $ be some basis in which $ x $ is $ k $-sparse. We use $ \boldsymbol{\phi} $ to mean the basis and it's matrix representation interchangeably. Let $ \{y_k\} $ be a set of $ m  $ measurements of $ x $ in some basis $ \boldsymbol{\psi} $, and let 
\begin{equation}\label{min_measurements}
m>\mu(\mathbf{\phi,\psi})^2 \cdot k\cdot \log\frac{n}{\delta}
\end{equation}
Then with probability greater than $ 1-\delta $:
\begin{equation}\label{reconstruction}
	x=\arg\min_{\mathbf{x}' \in \mathbb{R}^n} \| \mathbf{x}' \|_1 \quad \text{s.t.} \quad y_k=(\phi \mathbf{x}')_k
\end{equation}
This implies that if we take $ m \ge k \log^2 n $ measurement of $ x $ in some random basis, we can expect to reconstruct $ x $ with high probability.\\ 
Practically, $ m \approx k \log n$ measurements are usually sufficient to reconstruct the signal exactly, and signals can be approximately reconstructed from as few as $ m=O(k) $ measurements. \\
The measurements used in this paper are random elements of the Dragon Wavelet group\cite{feldman2016power}, which can be computed efficiently (in $ O(n \log n) $ time) and in-place for optimal memory use. The Dragon wavelet group resemble fractal noise patterns, and so are highly incoherent with natural images as well as other signals we would be likely to measure. 

\subsection{Comparison scenarios}
Next we will present the ideas behind a number of reconstruction methods, a short discussion of how they work and a quantitative comparison between them. This is done for three scenarios: 
\begin{itemize}
	\item Algorithms for reconstruction under the assumption of $ L_1 $ sparsity: This is the original premise of compressive sensing, however it is only applicable for signals that are already $ L_1 $ sparse. Natural images are not sparse in the $ L_1 $ sense, but usually have a sparse representation in other basis such as Discrete Cosine Transform (DCT - used in JPEG compression) or various wavelets (such as Haar wavelets, used for JPEG2000 compression). This leads to the proposal to "sparsify" an image and then use compressive sensing; a proposal which is discussed in section \ref{L1_methods}.
	\item Algorithms for reconstruction under Total Variation (TV) sparsity constraints: Total variation is an over-complete base in which natural images are sparse or nearly sparse (in essence, total variation is the $ L_1 $ norm of the discrete gradient of an image). While the standard proof of correct reconstruction can not be applied, the argument from incoherence holds. In general, algorithms that do not depend directly on the orthogonality of the measurement basis can also be easily modified to take advantage of any sparsifying norm or psudo-norm. 
	\item Reconstruction under noise: Most of the literature on compressive sensing focuses on the pure theoretical problem, where measurements in the compressive base are noiseless. This approach is easy to simulate, but does not cover the use-case where compressive sensing is most appropriate: noisy environments where a complete set of measurements is not available or not practical. A prime example is low-light imaging, where measurement noise is inversely proportional to the time over which averaging is done. In these scenarios, the improvement in acquisition time provided by compressive sensing would be negated by the increased time for each measurement required to achieve low noise. 
\end{itemize}

\section{$ L_1 $ Norm minimizing algorithms}
\label{L1_methods}
In order for compressive sensing minimizing the $ L_1 $ norm of a signal to work, the signal we are reconstructing must be sparse in the $ L_1 $ norm. This is not true for natural images, although the success of image compression algorithms based on re-sampling (such as JPEG and JPEG2000) indicates that there exists a basis in which most natural images are sparse. Abusing notation as we did in the theoretical background, let $ \boldsymbol{\psi} $ be our sampling basis as a matrix and $ \boldsymbol{\phi} $ be a sparsifying basis (such as DCT) as a matrix. We now create a new reconstruction problem, denoting $ \mathbf{z} = \boldsymbol{\phi} \mathbf{x} $, the image in the sparsifying base, so that $ y_k = (\psi \mathbf{z})_k = (\psi \phi \mathbf{x})_k $ are our measurements. The optimization problem becomes: 
\begin{align}
	\mathbf{x} = \phi^{-1} \mathbf{z}, \qquad \mathbf{z} = \arg\min_{\mathbf{z}'\in \mathbb{R}} \| \mathbf{z}' \|_1 \qquad \text{s.t.} \qquad y_k = (\boldsymbol{\psi} \mathbf{z'})_k
\end{align}
However, we can not get direct access to $ \mathbf{z} $ without already having the original signal $ \mathbf{x} $. We would like to re-sample $ \mathbf{x} $ using $ \boldsymbol{\phi} $ and then apply the sparsifying base, however $ \boldsymbol{\psi} $ does not in general commute with $ \boldsymbol{\phi} $, so $ \boldsymbol{\psi} \mathbf{z} = \boldsymbol{\psi}\boldsymbol{\phi} \mathbf{x} \neq \boldsymbol{\phi}\boldsymbol{\psi} \mathbf{x}$. For the sake of simulation, we can ignore this fact and attempt to reconstruct $ \mathbf{z} $ by computing $ \mathbf{z} = \boldsymbol{\phi} \mathbf{x} $ and $ \mathbf{x} = \phi^{-1} \mathbf{z} $ as pre- and post-calculation steps, as is done in the comparison part of this section, but this is not applicable to a real-world system. \\
Another option (inspired by \cite{hiesenberg_rep}) is to move the sensing matrix (rather than the signal) in to the sparsifying basis. This takes the form: 
\begin{gather}\label{sparsifying_Hisenberg}
	\tilde{\mathbf{y}} = \boldsymbol{\phi} \mathbf{y} \\
	\tilde{\boldsymbol{\psi}} = \boldsymbol{\phi}^{-1}\boldsymbol{\psi}\boldsymbol{\phi}
\end{gather}
Except that $ \boldsymbol{\psi} $ is not a square matrix. While mathematical tools exists to approximate $ \tilde{\boldsymbol{\psi}} $, they do not have a closed form and are not efficiently computable. \\
To conclude this section, we gives a short history of relevant optimization algorithms, focusing on the incremental improvements each one provides. However, $ L_1 $ minimization are not in general optimal for imaging applications and the algorithms here are brought merely for completeness. 
\subsection{Off-the-shelf algorithms}
First, a brief review of the most standard optimization algorithms, available as library functions in most scientific and programming environments. Many of these algorithms are packaged together as a single library function, such as the simplex method, interior-point and ellipsoid algorithm are in MATLAB. The disadvantages of these methods are their slow convergence rate and prohibitive memory requirements, which make them unsuitable for large problems. Additionally, these methods have little to no tolerance for noisy measurements – when applied to imperfect data, they often fail to converge. We therefore start our more in-depth review with more advance algorithms. \\
\subsubsection{Orthogonal Matching Pursuit (OMP)} 
OMP is conceptually based on building a dictionary of simple image elements, and recovering a new image by combining the best matching elements of the dictionary \cite{tropp2007signal,tropp2005signal}. An iterative process finds the best matching element in the dictionary and adds it to the reconstruction until the residual measurement is sufficiently small. \\
This method is often proposed for compressive sensing reconstruction, though it is not entirely appropriate: firstly, for a dictionary lookup method to work well the dictionary should be over-complete, which means we must take many more samples than indicated by the information bound. Secondly, the reconstruction guaranty is not independent of the signal as it is in linear programming. For these reasons, OMP is more suited for image de-noising, where it is often used. 
\subsubsection{Ridge Regression, aka LASSO} 
LASSO is a small alteration of standard linear optimization problems that compensates for noisy measurements\cite{candes2008introduction,howland2014compressive}.  Formally, the problem is defined as:
\begin{equation}\label{LASSO}
\mathbf{ x }=\min_{\mathbf{ x }'\in\mathbb{R}^n}\| \mathbf{ y } - \boldsymbol{\psi}\mathbf{ x }' \|_2 \quad \text{s.t.} \quad \| \mathbf{ x }'\|_1 \leq \tau  
\end{equation}
Where $ \tau $ is the assume $ L_1 $ bound of the signal. In principle this can be seen as searching for a signal with a small $ L_1 $ norm that matches the measurements with minimal noise. The current state of the art LASSO algorithms are significantly faster and use less memory than standard linear programming and OMP, but memory usage scales prohibitively with image size, with mega-pixel images well beyond the resources of widely available computing systems. 
\subsection{Gradient projection for sparse reconstruction (GPSR)} GPSR is a newer algorithm developed specially for compressed sensing \cite{howland2014compressive,nowak2007gradient}. This approach formulates the problem as a bound constraint quadratic program. By defining $ \mathbf{ x } = \mathbf{u} - \mathbf{v} \quad \mathbf{u} \geq 0, \mathbf{v} \geq 0 $ we get $ \| x\|_1 = \mathbf{u} + \mathbf{v} $, and we can recast the optimization problem to the form:
\begin{align}\label{BQPB}
\min_{\mathbf{u,v}}  \frac{1}{2} \| \mathbf{y}- A(\mathbf{u} - \mathbf{v}) \|_2^2 + \lambda\sum (\mathbf{u} + \mathbf{v}) \qquad\text{s.t.} \qquad& \mathbf{u} \geq 0 \\ \nonumber
& \mathbf{v} \geq 0
\end{align}
The GPSR algorithm now performs minimization step by searching along the negative gradient of the loss function projected into the non-negative orthant for a point that is within the problem bounds. There is also a variation of the algorithm based on the approach of Barzilai and Borwein\cite{barzilai1988two}, which does not guaranty a decrease in the loss function every step but tends to converge faster. \\

\subsection{Spectral Projected Gradient for $ L_1 $ minimization (SPGL1)} \label{SPGL1} SPGL1 is based on the observation that in the case of, e.g., LASSO formulation, if $ \tau $ is a controllable parameter, the expression to minimize defines a trade-off curve between the $ L_1 $ norm of the signal and the least-squares fit to the measurements. The crucial realization for this algorithm relies on introducing another formulation of the problem, Basis Pursuit De-Noising (BPDN): 
\begin{equation}\label{BPDN}
	\mathbf{ x }=\min_{\mathbf{ x }'\in\mathbb{R}^n}\| \mathbf{ x }'\|_1  \quad \text{s.t.} \quad \| \mathbf{ y } - \boldsymbol{\psi}\mathbf{ x }' \|_2\leq \sigma  
\end{equation}
It can now be shown that the dual solution to eq. \ref{LASSO} yields information on how to update the parameter $ \tau $ to obtain a solution much closer to the one given by \ref{BPDN}. These two solutions meet on the desired trade-off curve, giving an optimal solution for the chosen values of $ \tau $ and $ \sigma $.\\
\subsubsection{Theoretical conclusions} GPSR and SPGL1 both present a striking advantage over their predecessors: the iterative step only requires vector-matrix products with $ A $ and $ A^T $, making them efficient in memory and computation time. This property can be utilized farther by providing measurement matrices that allow for fast vector products via recursive structure, such as Hadamard and Fourier transform. The algorithms introduced in the next section also have this advantage. \\

\subsection{Practical comparison}
\begin{figure}[h]
	\centering
	\includegraphics[width=0.7\textwidth]{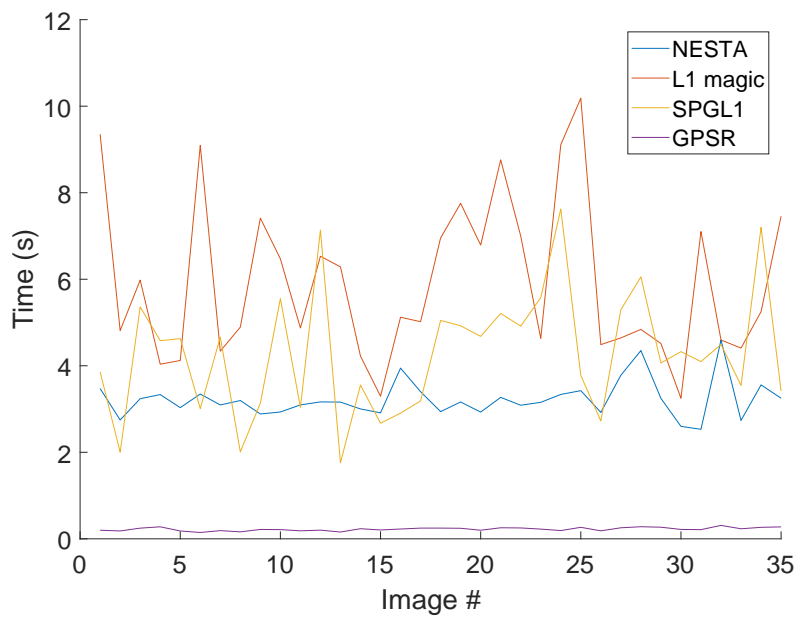}
	\caption{Comparison of runtime for the $ L_1 $ algorithms presented. }
	\label{l1_times}
\end{figure}

\begin{figure}[h]
	\centering
	\includegraphics[width=0.7\textwidth]{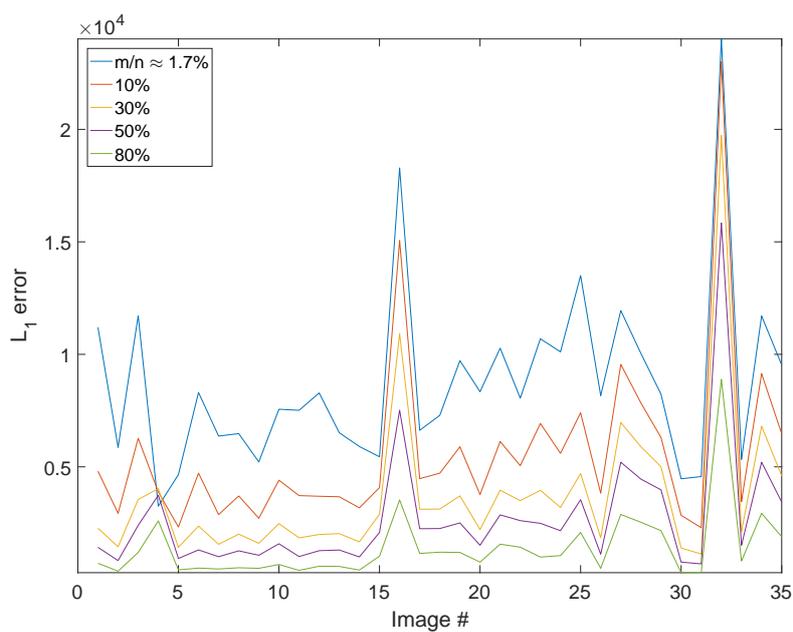}
	\caption{$ L_1 $ error for different images, for different numbers of measurements. }
	\label{l1_quality}
\end{figure}

The following performance comparison is provided for completeness: GPSR and SPGL1 (described here) are compared with L1 Magic and NESTA (described in the next section) for a variety of images, benchmarked for speed (Fig. \ref{l1_times}). The times shown are for a compression ration of 80\%, where reconstruction time increases with the number of measurements. Code profiling shows that there are two causes for the this: firstly, longer measurement vectors directly affect calculation time because more operations are needed to process them. Secondly, with less constraint information the optimum described by the measurements is reached in fewer iterations and convergence stops. \\
Reconstruction quality is identical between the algorithms for the $ L_1 $ case (Fig. \ref{l1_quality}), and one example is brought for illustration (Fig. \ref{l1_example}). All the images used for comparison can be found in the appendix. 
\begin{figure}[H]
	\centering
	\begin{subfigure}[t]{0.6\textwidth}
		\includegraphics[width=1\textwidth]{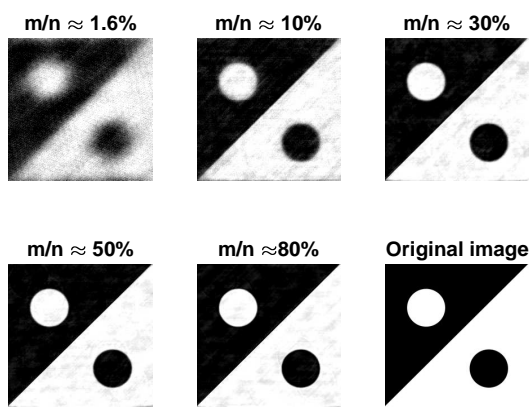}
		\caption{Image \#1, reconstructed using NESTA}
	\end{subfigure}
	~
	\begin{subfigure}[t]{0.6\textwidth}
		\includegraphics[width=1\textwidth]{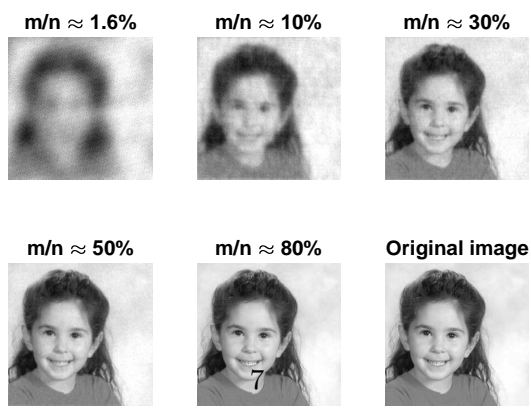}
		\caption{Image \#7, reconstructed using NESTA}
	\end{subfigure}
	\caption{Sample images. Even using $ 80\% $ of the measurements, the reconstruction is imperfect. While the reasons for this are beyond the scope of this paper, some ideas are brought in section \ref{conclutions}. }
	\label{l1_example}
\end{figure}

\section{Algorithms capable of total variation minimization }

\subsection{L1 Magic}
The L1 Magic package is not a single algorithm but several, developed together as a educational tool to prove computational tractability of compressive sensing and reconstruction from incomplete measurements. While these algorithms are not optimized nor fully robust, they are the package of choice for many educators in the field of compressive sensing. The package can be divided in to two primary parts: a set of $ L_1 $ solvers using a primal-dual approach for linear problems (similar in concept to \ref{SPGL1}), and a $ TV $ pseudo-norm solver using a log-barrier algorithm, formulated as second-order cone problems. The algorithms within each part differ only by the Newton step used to generate the gradient followed in each iteration until convergence is reached. \\
This package implements standard algorithms remarkably well and is very suitable for educational purposes thanks to it's simple structure. However, lack of optimization and robustness make it in-optimal for large scale problems. Specifically, both the $ L_1 $ and $ TV $ implementations fail entirely for ill-posed problems, such as sampling near the information content of the image (as described in section \ref{recon_theory}) or measurements with noise, while other algorithms show varying degrees of success (section \ref{noisy_recon}). 

\subsection{Nestorov's algorithm (NESTA)}
The NESTA algorithm is based on a method developed by Yurii Nesterov for deriving optimization algorithms that achieve a convergence rate he proved was optimal two decades earlier \cite{nesterov1983method}. The method combines smoothing with an augmented gradient to achieve a $ \epsilon = O(\frac{1}{k^2}) $ convergence rate (where $ \epsilon $ is the residual and $ k $ is the number of steps) \cite{becker2011nesta}. \\
Nestorov's algorithm can be applied to any smooth, convex loss functions over a convex feasible set. The function's derivative is assumed to be Lipschitz, and the algorithm then iteratively estimates three sequences: \\
$ {x_k} $: the minimum of the loss function, \\
$ {y_k} $: a Lipschitz bounded next guess for $ x_k $, and \\
$ {z_k} $: a smoothed gradient step, which takes into account all the previous steps and a proximity function that enforces the convex set's bounds (for constrained problems). \\
NESTA is the only algorithm presented in this paper with a better theoretical convergence rate than $ O(\frac{1}{k}) $, and it is capable of both $ L_1 $ and total variation minimization. 

\subsection{TVAL3}
The TVAL3 algorithm is base on an augmented Lagrangian multiplier approach, using an alternating direction non-monotone line search to minimize an augmented Lagrangian type loss function \cite{howland2014compressive,li2013efficient,yang2011alternating}. TVAL3 can solve any optimization problem of the form:
\begin{equation}\label{TVAL3eqn}
\min_{x,y} f(x,y) \qquad \text{s.t.} \quad h(x,y) = 0
\end{equation}
Under the assumption that $ h $ is differentiable with respect to $ x $ and $ y $ and $ f $ is differentiable with respect to $ x $ (but not necessarily in respect to $ y $). In our case, we use $ f = \|x\|_{TV} $ and $ h = \mathbf{ y } - \boldsymbol{\psi} \mathbf{x} $. The problem is then approximated by the following augmented Lagrangian form:
\begin{equation}\label{Lagrangian}
\mathcal{L}(x,y,\lambda; \mu) = f(x,y) - \lambda^T h(x,y) + \frac{\mu}{2} h(x,y)^T h(x,y), \quad \mu > 0
\end{equation}
which can be minimized iteratively under the assumption of uneven complexity: minimizing with respect to $ y $ is of much lower complexity than minimizing with respect to $ x $. \\
TVAL3 alternates between minimizing the augmented Lagrangian with the current Lagrange multipliers and updating the Lagrange multipliers and $ \mu $, until sufficient convergence is reached. \\
While difficult to compare analytically to other algorithms on account of the alternation between the primal and dual problems, TVAL3 is remarkably fast. The algorithms inability to optimize $ L_1 $ problems is not a shortcoming for imaging applications, as noted in the beginning of section \ref{L1_methods}.

\subsection{Practical comparison}
\begin{figure}[H]
	\centering
	\includegraphics[width=0.7\textwidth]{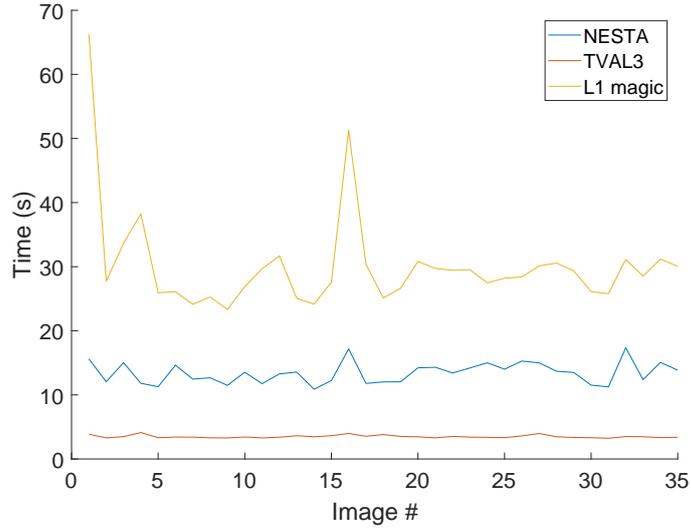}
	\caption{Comparison of runtime for the $ TV $ algorithms presented. }
	\label{tv_times}
\end{figure}

\begin{figure}[H]
	\centering
	\begin{subfigure}[t]{0.3\textwidth}
		\includegraphics[width=1\textwidth]{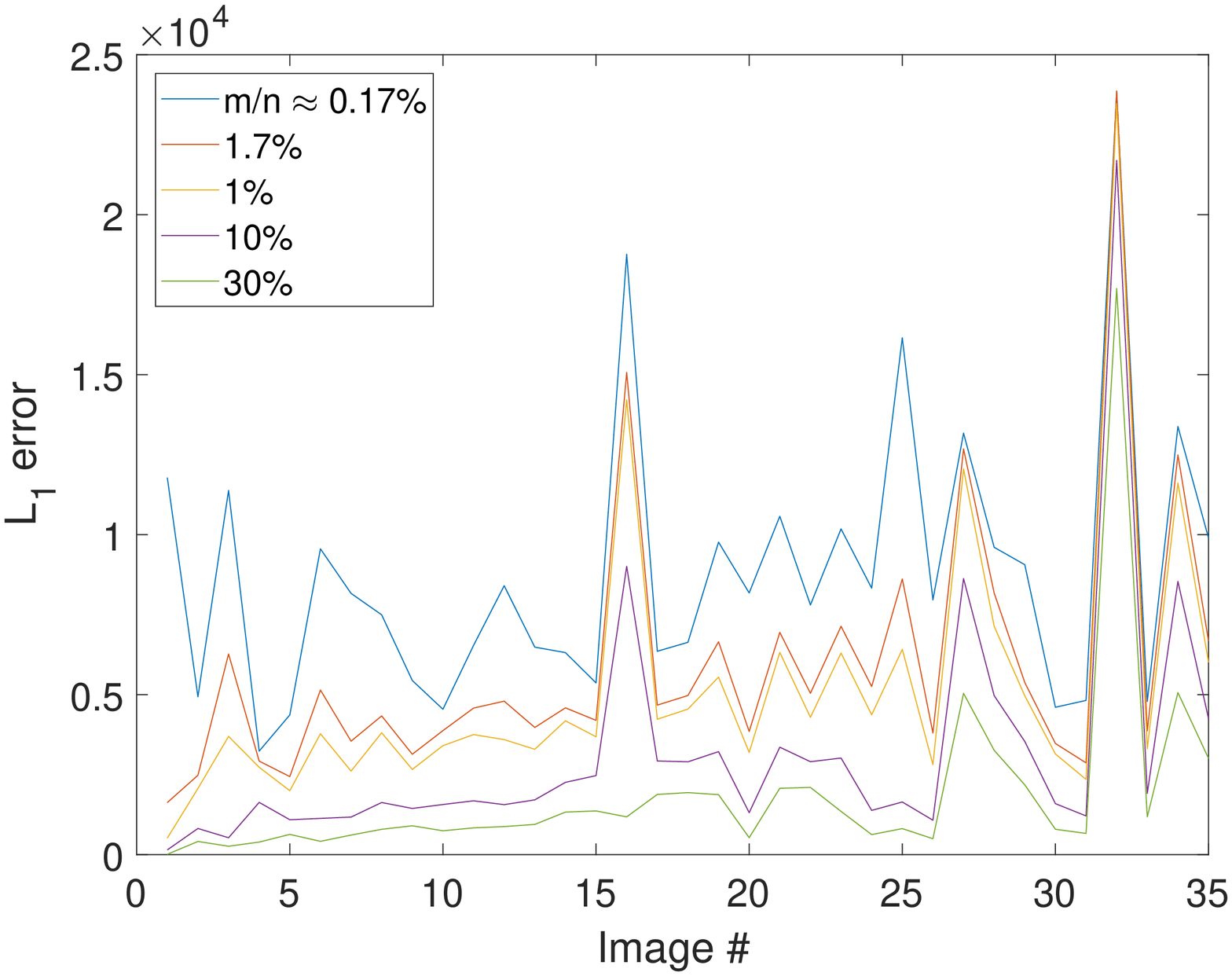}
		\caption{$ L_1 $ Magic}
	\end{subfigure}
	~
	\begin{subfigure}[t]{0.3\textwidth}
		\includegraphics[width=1\textwidth]{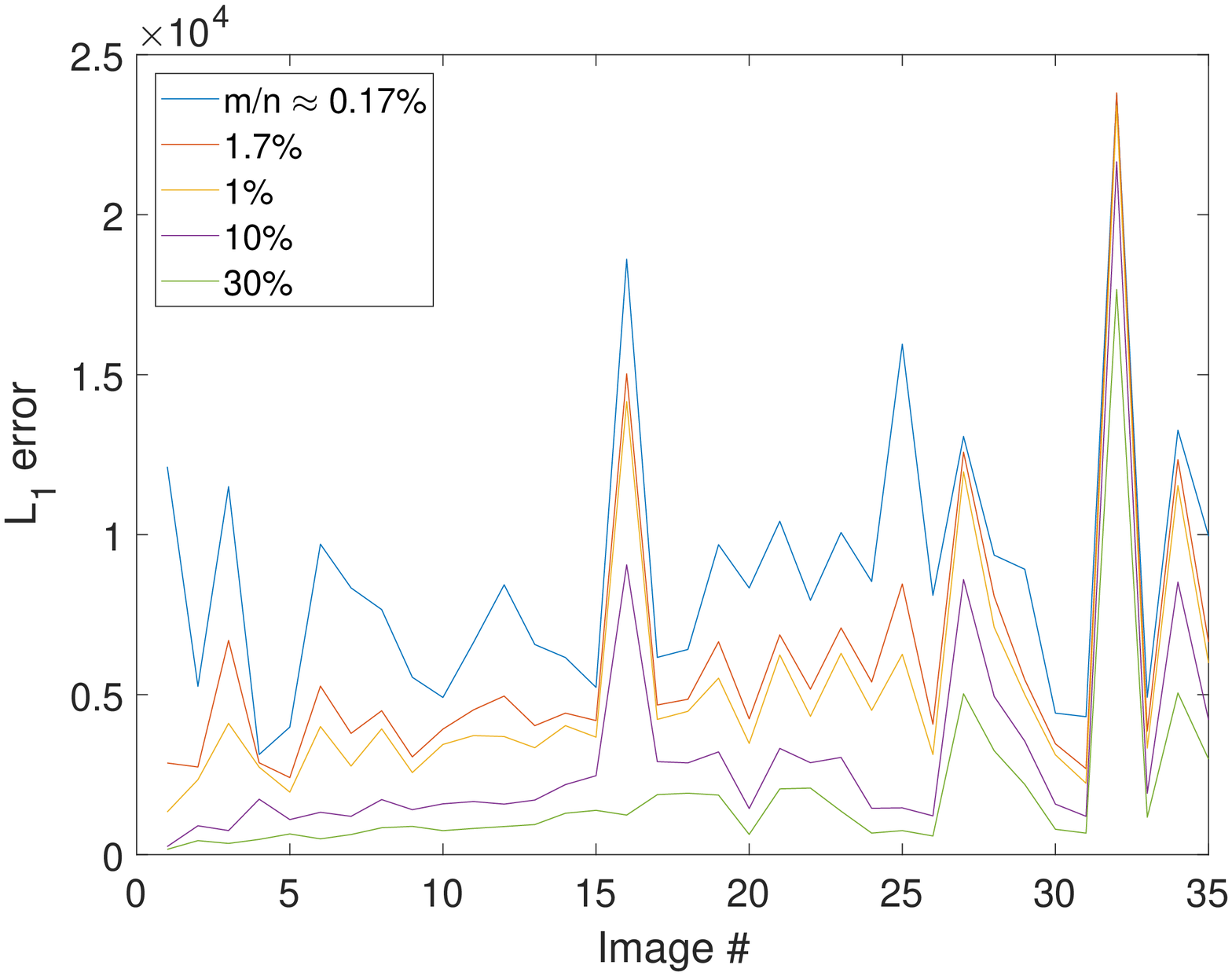}
		\caption{NESTA}
	\end{subfigure}
	~
	\begin{subfigure}[t]{0.3\textwidth}
		\includegraphics[width=1\textwidth]{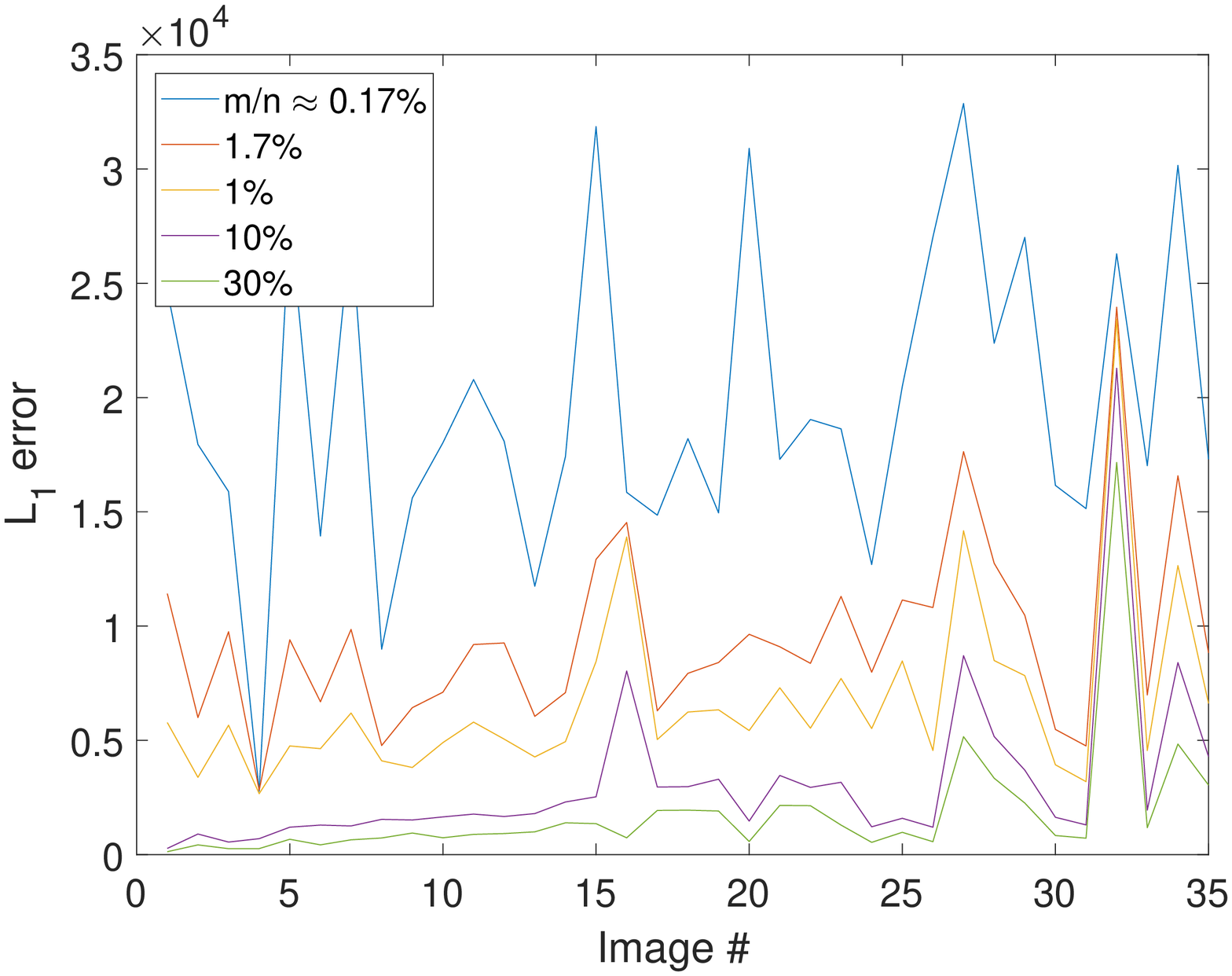}
		\caption{TVAL3}
	\end{subfigure}
	\caption{$ L_1 $ error for different images, for different numbers of measurements. Here we find that while the algorithms show identical results for well-posed problems, as the number of measurements decreases TVAL3 is less able to reconstruct the image. This will become more of an issue with noisy measurements, as shown in section \ref{noisy_recon}. Note that while the $ L_1 $ error does decrease when moving from $ 10\% $ of the measurements to $ 30\% $, the difference is less pronounced. This effect can be seen clearly in the example images (Fig. \ref{tv_examples}). }
	\label{tv_quality}
\end{figure}

\begin{figure}[H]
	\centering
	\begin{subfigure}[t]{0.7\textwidth}
		\includegraphics[width=1\textwidth]{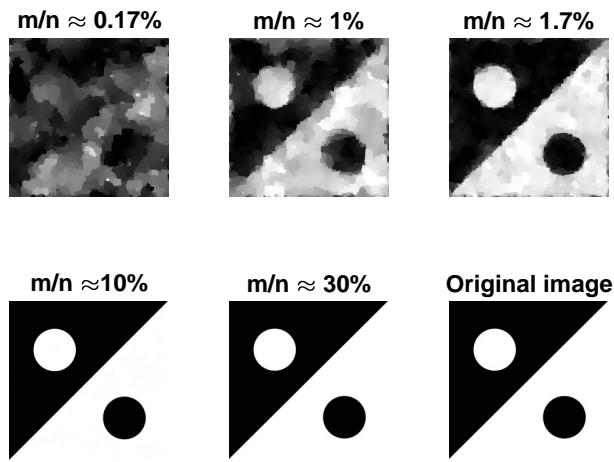}
		\caption{Image \#1, reconstructed using NESTA}
	\end{subfigure}
	~
	\begin{subfigure}[t]{0.7\textwidth}
		\includegraphics[width=1\textwidth]{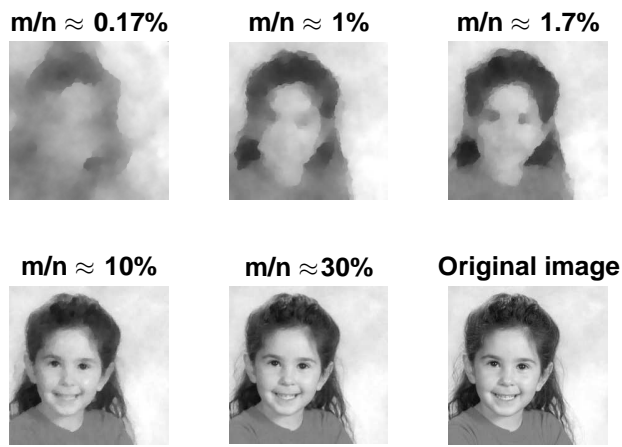}
		\caption{Image \#7, reconstructed using NESTA}
	\end{subfigure}
	\caption{Sample images. The visual quality is greatly improved over $ L_1 $ reconstruction with a fraction of the measurements, while the $ L_1 $ error is comparable. Some thoughts on this difference can be found in section \ref{conclutions}. }
	\label{tv_examples}
\end{figure}

The comparisons between algorithms for total variation minimization are shown in figures \ref{tv_times}, \ref{tv_quality} and \ref{tv_examples}. TVAL is nearly three times as fast as NESTA and ten times faster than L1 Magic, so it is the obvious choice for well-conditioned problems. TVAL's shortcoming lies in situations where the number of measurements is close to the information limit or when there is significant measurement noise (as described in section \ref{noisy_recon}), where the quality of reconstruction decreases rapidly. \\
\subsubsection{Comparison to $ L_1 $ algorithms}  The most striking difference between $ L_1 $ and total variation algorithms is the number of measurements required to achieve good quality reconstruction. While the values for $ L_1 $ error are similar in the examples brought here, the total variation algorithms need a smaller percentage of the the measurements, and the reconstructed image is more similar the original even when the $ L_1 $ error indicates the reconstruction qualities are comparable. The time required to reconstruct an image also shows consistent variation between the algorithm classes: Total variation algorithms take approximately three times longer to reconstruct an image than $ L_1 $ methods, and profiling shows that nearly all the difference is due to calculating the total variation pseudo-norm, while the number of iterations is almost identical (between algorithms of similar structure). 


\section{Reconstruction under noise}
\label{noisy_recon}
Finally, we would like to present the effects of measurement noise on reconstruction quality, to complete the preparation for using compressive sensing in real-world systems. The modele used here is white Gaussian additive noise with mean $ 0 $ and standard deviation chosen to provide the desired SNR. While the low SNR value of $ 3 dB $ might seem extreme, it was chosen to provide similar conditions to experimental data sets such as the ones in \cite{sher2018low}. The number of measurements was chosen to be $ 8\% $ of the number of pixels for the same reason. \\
An interesting feature of noisy reconstruction not shown in the graphs is that reconstruction takes longer for higher SNR values. Here the length of the measurement vectors is fixed, so we conclude that without enough information to reconstruct the image correctly the algorithm "gives up" early. The relative times between different algorithms are similar to the ones without noise. 

\begin{figure}[H]
	\centering
	\begin{subfigure}[t]{0.4\textwidth}
		\includegraphics[width=1\textwidth]{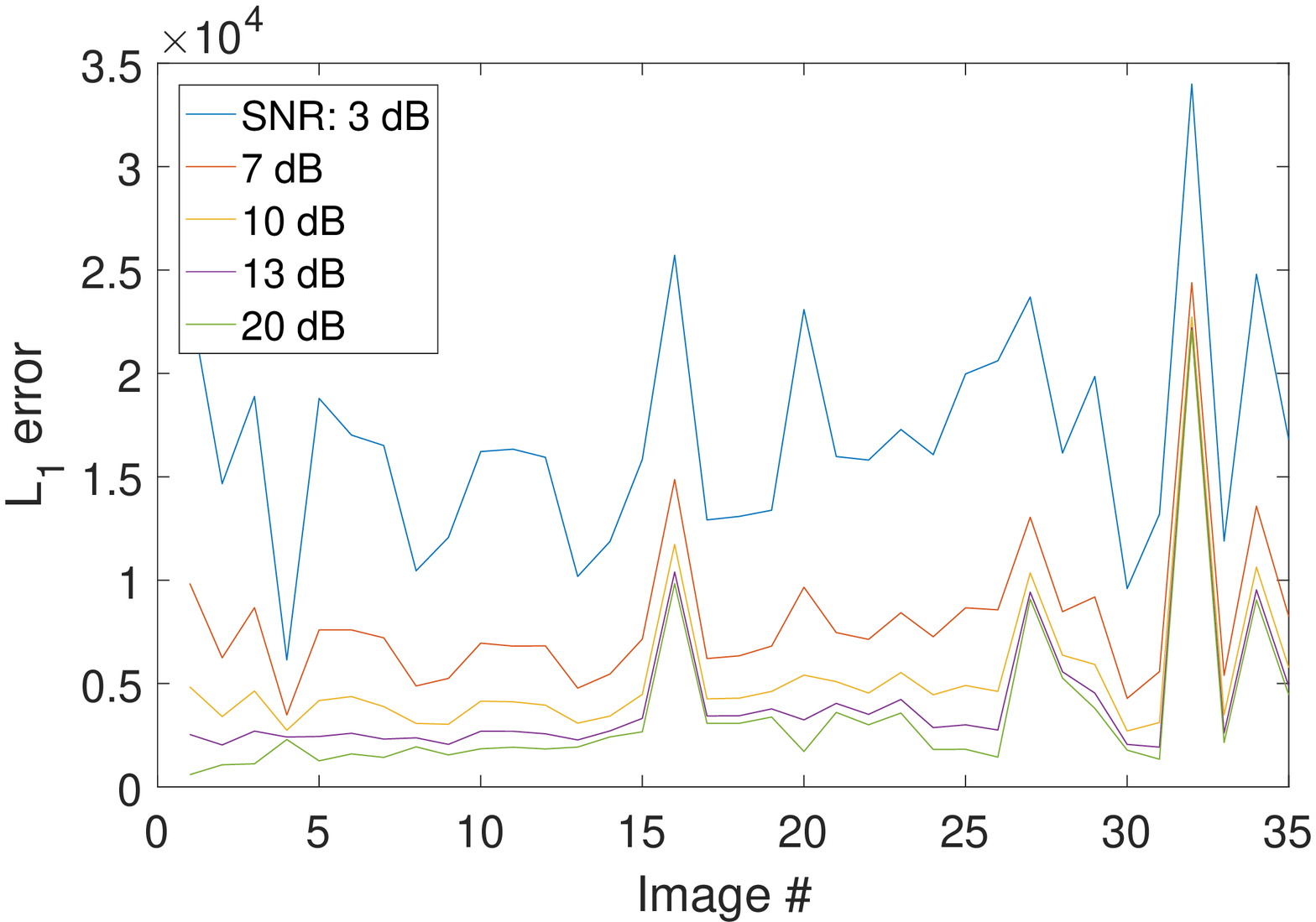}
		\caption{NESTA}
	\end{subfigure}
	~
	\begin{subfigure}[t]{0.4\textwidth}
		\includegraphics[width=1\textwidth]{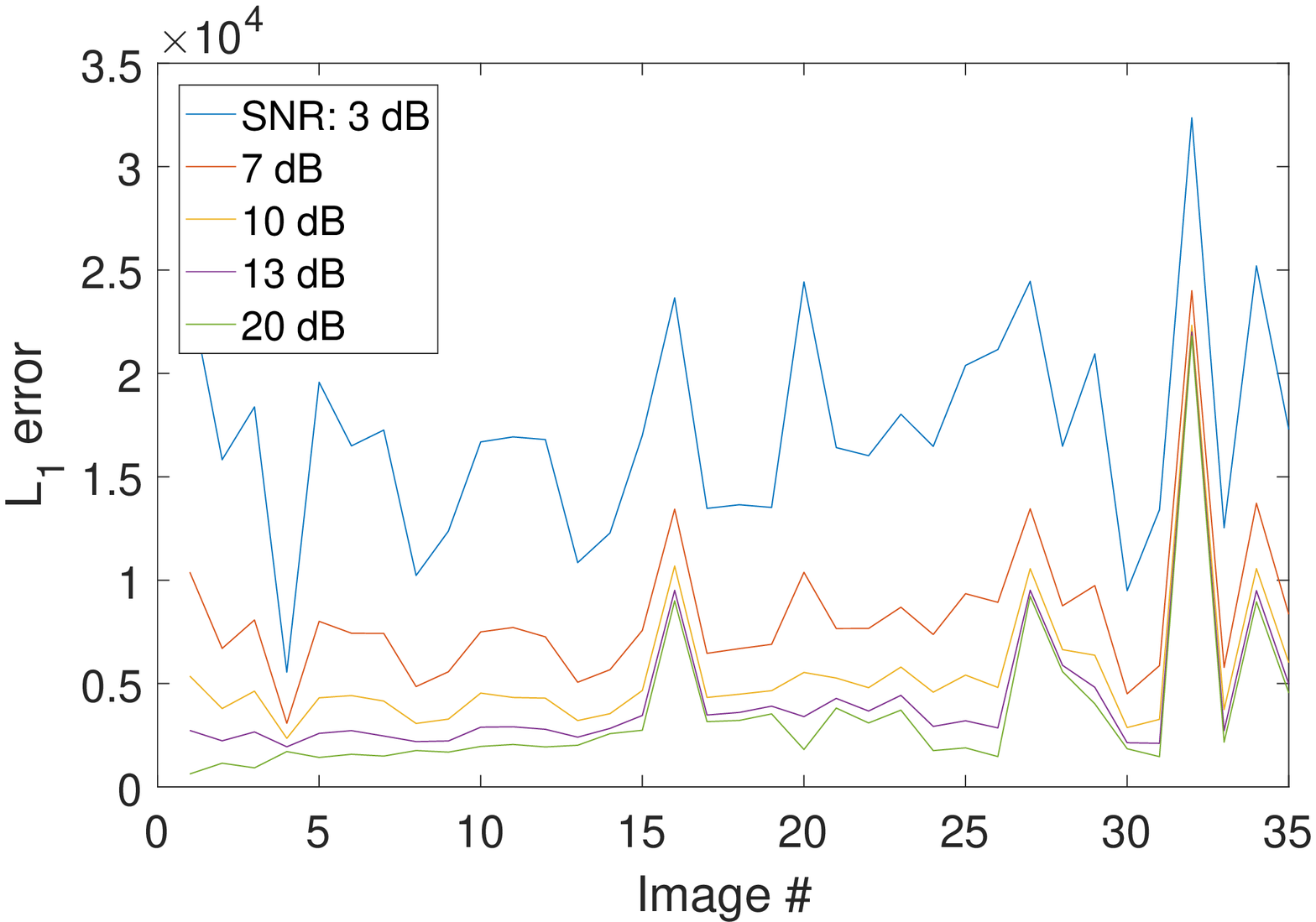}
		\caption{TVAL3}
	\end{subfigure}
	\caption{$ L_1 $ error for different images, for different SNR. There is a pronounced difference between NESTA and TVAL3 at very low SNR: while NESTA reconstructs an image that maintains the overall features and can be improved using post-processing tools, TVAL3 fails completely. This is crucial to experimental systems, and is illustrated in the example images (Fig. \ref{noise_examples}). }
	\label{noise_quality}
\end{figure}
\begin{figure}[H]
	\centering
	\begin{subfigure}[t]{0.7\textwidth}
		\includegraphics[width=1\textwidth]{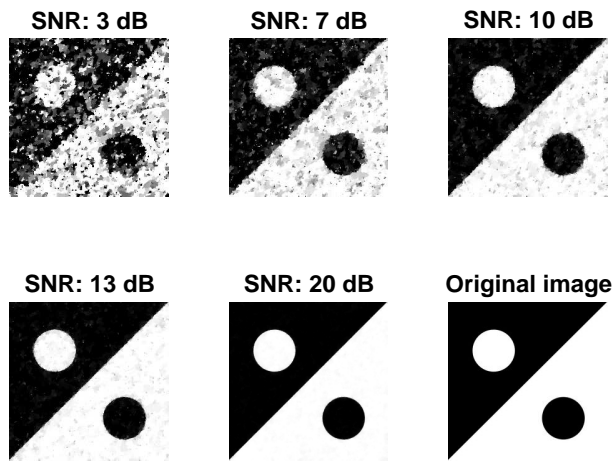}
		\caption{Image \#1, reconstructed using NESTA}
	\end{subfigure}
	~
	\begin{subfigure}[t]{0.7\textwidth}
		\includegraphics[width=1\textwidth]{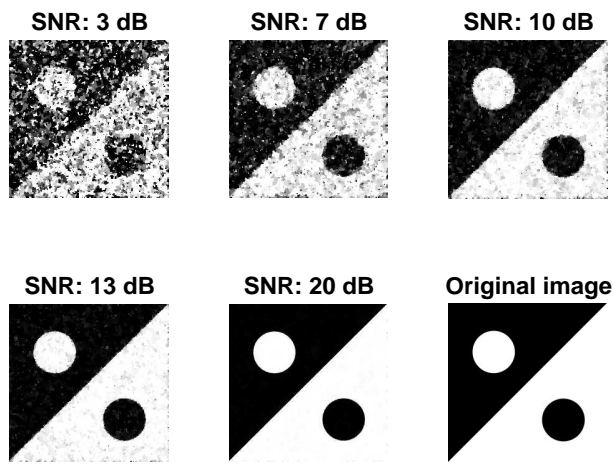}
		\caption{Image \#1, reconstructed using TVAL3}
	\end{subfigure}
	\caption{Image 1, reconstructed using different algorithms. At high SNR both algorithms preform well. At low SNR, both algorithms preform poorly (as expected), but NESTA better maintains the major features allowing for improvement via post-processing. }
	\label{noise_examples}
\end{figure}

\section{Summary}
\label{conclutions}
Our first conclusion is that total variation minimization is more appropriate for natural images than $ L_1 $ minimization under a sparsifying basis. Beyond the experimental data, we justify this with the following intuition: In both cases (TV and $ L_1 $ under a wavelet transform) the vector we reconstruct is not mathematically sparse; rather, many of it's entries are very small, and the reconstruction takes them to be zero. Since wavelet transforms are global operations, this leaves interference patterns that require more measurements to eliminate. Additionally, under total variation minimizations the zeroed elements are local, leading to image distortion that have smaller $ L_1 $ distance and are less noticeable to the eye. \\
Of the algorithms described here, two stand out: NESTA and TVAL3. Both are fast and efficient, and both use only matrix-vector products to leverage efficiently calculable basis (such as Hadamard or Dragon wavelets). This difference is especially significant as the problem scales to mega-pixel images, where speed and memory requirements become prevalent concerns. Both algorithms are also robust: their performance is not dependent on the exact values of the tunable input parameters and is only weakly dependent the image provided. For experimental uses the trade-off between noise tolerance and speed is significant, and will ultimately decide which algorithm is appropriate on a application specific basis. \\

\section{Future work}
A major point left out in this work is comparison of different sampling bases, such as can be found in \cite{wang2010variable}. The choice of Dragon noiselets is based on ease of use, and as such was never fully quantified. Even within the family of self-similar noiselets, scrambled Hadamard sampling vectors present superior memory performance and are nearly as easy to implement. Additionally, we do not discuss parallelization or other recent optimization techniques, which would surely play a part in any industrial-scale application. \\
Finally, while this paper was written using insights gained from real-world data, it's scope does not extend far enough to contain a quantitative discussion of reconstruction of images outside the virtual realm. 

\section{Acknowledgments}
We would like to thank Yair Weiss for the background and guidance enabling this work, and Hagai Eisenberg from the Racah Institute of Physics for providing the framework and resources to apply these conclusions to real data.

\newpage
\appendix
\section{Images used for comparison}
\begin{figure}[H]
\centering
\begin{minipage}{0.25\textwidth}
	\includegraphics[width=1\textwidth]{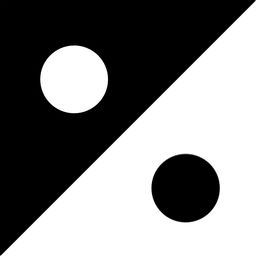}
	\caption{Image \#1}
\end{minipage} \qquad\qquad
\begin{minipage}{0.25\textwidth}
	\includegraphics[width=1\textwidth]{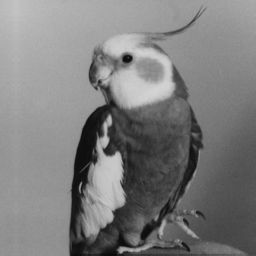}
	\caption{Image \#2}
\end{minipage} \qquad\qquad
\begin{minipage}{0.25\textwidth}
	\includegraphics[width=1\textwidth]{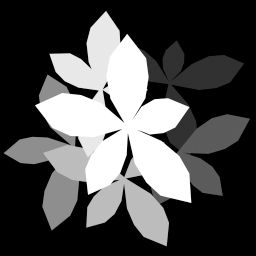}
	\caption{Image \#3}
\end{minipage} \qquad\qquad
\begin{minipage}{0.25\textwidth}
	\includegraphics[width=1\textwidth]{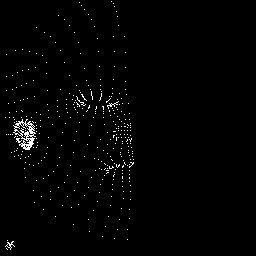}
	\caption{Image \#4}
\end{minipage} \qquad\qquad
\begin{minipage}{0.25\textwidth}
	\includegraphics[width=1\textwidth]{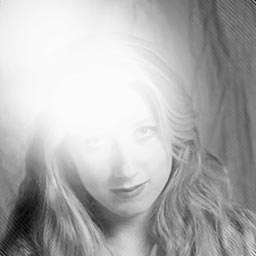}
	\caption{Image \#5}
\end{minipage} \qquad\qquad
\begin{minipage}{0.25\textwidth}
	\includegraphics[width=1\textwidth]{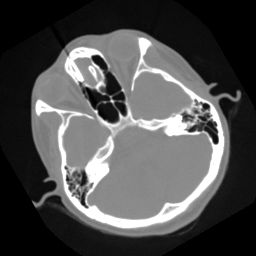} \label{image16}
	\caption{Image \#6}
\end{minipage} \qquad\qquad
\begin{minipage}{0.25\textwidth}
	\includegraphics[width=1\textwidth]{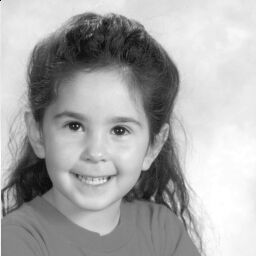}
	\caption{Image \#7}
\end{minipage} \qquad\qquad
\begin{minipage}{0.25\textwidth}
	\includegraphics[width=1\textwidth]{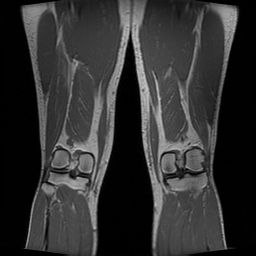}
	\caption{Image \#8}
\end{minipage} \qquad\qquad
\begin{minipage}{0.25\textwidth}
	\includegraphics[width=1\textwidth]{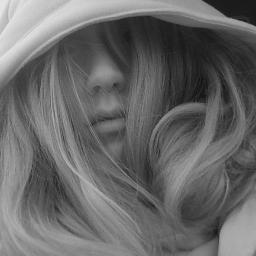}
	\caption{Image \#9}
\end{minipage} \qquad\qquad

\end{figure}

\begin{figure}[H]
\centering
\begin{minipage}{0.25\textwidth}
	\includegraphics[width=1\textwidth]{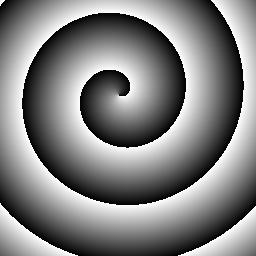}
	\caption{Image \#25}
\end{minipage} \qquad\qquad
\begin{minipage}{0.25\textwidth}
	\includegraphics[width=1\textwidth]{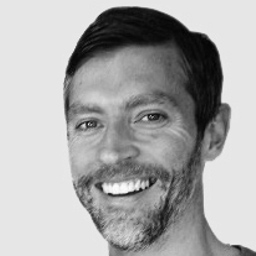}
	\caption{Image \#26}
\end{minipage} \qquad\qquad
\begin{minipage}{0.25\textwidth}
	\includegraphics[width=1\textwidth]{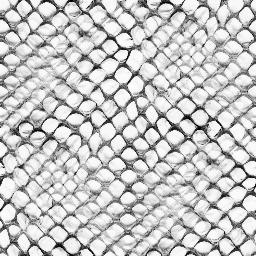}
	\caption{Image \#27}
\end{minipage} \qquad\qquad
\begin{minipage}{0.25\textwidth}
	\includegraphics[width=1\textwidth]{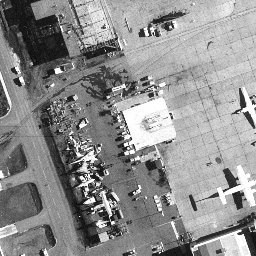}
	\caption{Image \#28}
\end{minipage} \qquad\qquad
\begin{minipage}{0.25\textwidth}
	\includegraphics[width=1\textwidth]{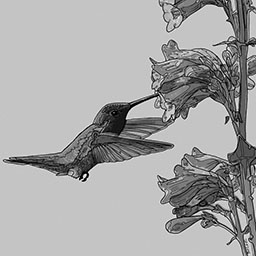}
	\caption{Image \#29}
\end{minipage} \qquad\qquad
\begin{minipage}{0.25\textwidth}
	\includegraphics[width=1\textwidth]{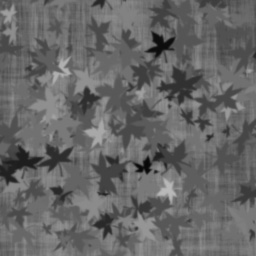}
	\caption{Image \#30}
\end{minipage} \qquad\qquad
\begin{minipage}{0.25\textwidth}
	\includegraphics[width=1\textwidth]{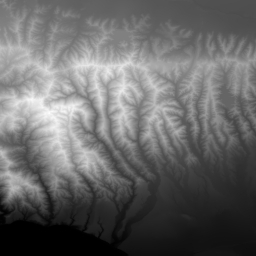}
	\caption{Image \#31}
\end{minipage} \qquad\qquad
\begin{minipage}{0.25\textwidth}
	\includegraphics[width=1\textwidth]{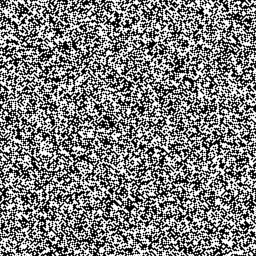}\label{image32}
	\caption{Image \#32}
\end{minipage} \qquad\qquad
\begin{minipage}{0.25\textwidth}
	\includegraphics[width=1\textwidth]{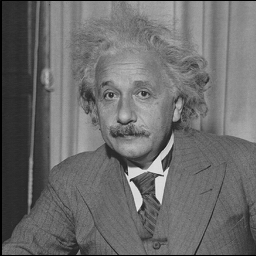}
	\caption{Image \#33}
\end{minipage} \qquad\qquad
\begin{minipage}{0.25\textwidth}
	\includegraphics[width=1\textwidth]{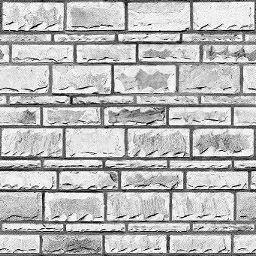}
	\caption{Image \#34}
\end{minipage} \qquad\qquad
\begin{minipage}{0.25\textwidth}
	\includegraphics[width=1\textwidth]{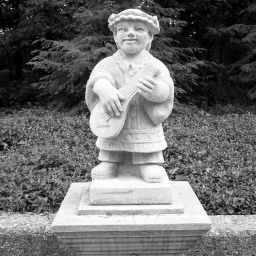}
	\caption{Image \#35}
\end{minipage} \qquad\qquad

\end{figure}

All the images used in the paper are included here, with the identifying numbering used above. Some features of the reconstruction are easy to see, e.g. image 32 (Figure \ref{image32}) is random noise and the reconstruction error is high, as expected. Image 16 (Figure \ref{image16}) also has high information content in the TV sense because of it's sharp edges. The code used for this paper is available on the author's website. 

\begin{figure}[H]

\centering
\begin{minipage}{0.25\textwidth}
	\includegraphics[width=1\textwidth]{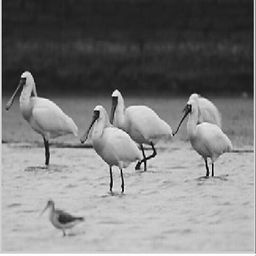}
	\caption{Image \#10}
\end{minipage} \qquad\qquad
\begin{minipage}{0.25\textwidth}
	\includegraphics[width=1\textwidth]{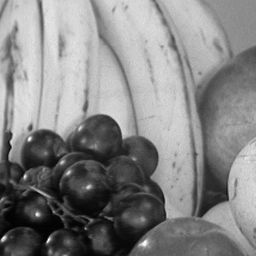}
	\caption{Image \#11}
\end{minipage} \qquad\qquad
\begin{minipage}{0.25\textwidth}
	\includegraphics[width=1\textwidth]{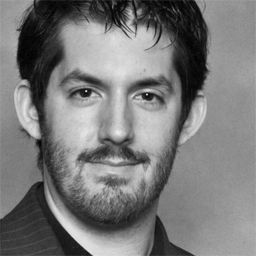}
	\caption{Image \#12}
\end{minipage} \qquad\qquad
\begin{minipage}{0.25\textwidth}
	\includegraphics[width=1\textwidth]{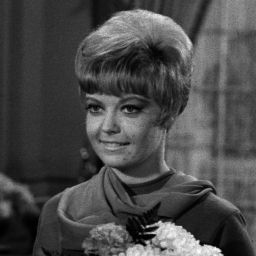}
	\caption{Image \#13}
\end{minipage} \qquad\qquad
\begin{minipage}{0.25\textwidth}
	\includegraphics[width=1\textwidth]{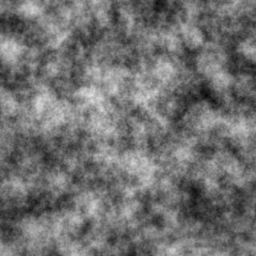}
	\caption{Image \#14}
\end{minipage} \qquad\qquad
\begin{minipage}{0.25\textwidth}
	\includegraphics[width=1\textwidth]{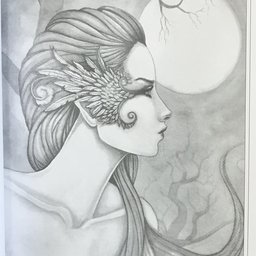}
	\caption{Image \#15}
\end{minipage} \qquad\qquad
\begin{minipage}{0.25\textwidth}
	\includegraphics[width=1\textwidth]{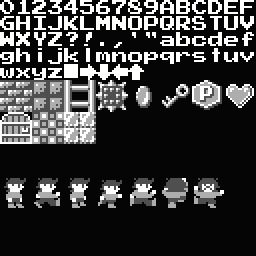}
	\caption{Image \#16}
\end{minipage} \qquad\qquad
\begin{minipage}{0.25\textwidth}
	\includegraphics[width=1\textwidth]{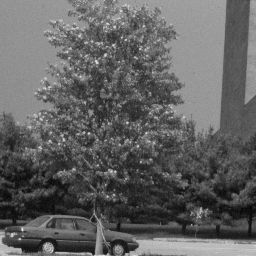}
	\caption{Image \#17}
\end{minipage} \qquad\qquad
\begin{minipage}{0.25\textwidth}
	\includegraphics[width=1\textwidth]{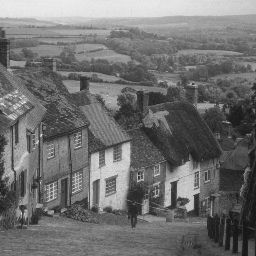}
	\caption{Image \#18}
\end{minipage} \qquad\qquad
\begin{minipage}{0.25\textwidth}
	\includegraphics[width=1\textwidth]{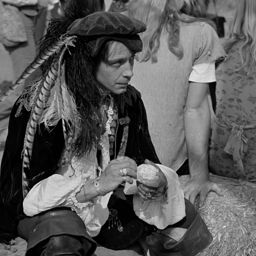}
	\caption{Image \#19}
\end{minipage} \qquad\qquad
\begin{minipage}{0.25\textwidth}
	\includegraphics[width=1\textwidth]{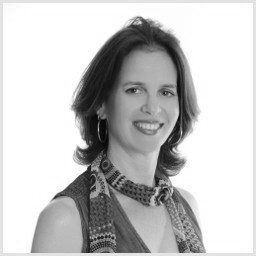}
	\caption{Image \#20}
\end{minipage} \qquad\qquad
\begin{minipage}{0.25\textwidth}
	\includegraphics[width=1\textwidth]{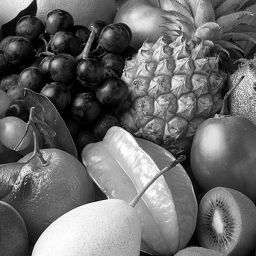}
	\caption{Image \#21}
\end{minipage} \qquad\qquad
\begin{minipage}{0.25\textwidth}
	\includegraphics[width=1\textwidth]{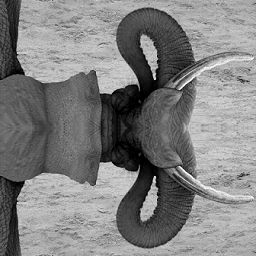}
	\caption{Image \#22}
\end{minipage} \qquad\qquad
\begin{minipage}{0.25\textwidth}
	\includegraphics[width=1\textwidth]{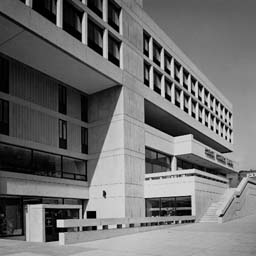}
	\caption{Image \#23}
\end{minipage} \qquad\qquad
\begin{minipage}{0.25\textwidth}
	\includegraphics[width=1\textwidth]{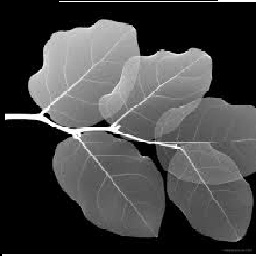}
	\caption{Image \#24}
\end{minipage} \qquad\qquad

\end{figure}

\end{document}